\documentclass{article}
\usepackage{FPSpro}
\usepackage[dvips]{graphicx}

\begin{document}
\title{KAON RARE DECAYS
  }
\author{
  Takeshi K. Komatsubara\\
  {\em KEK-IPNS, Tsukuba, Ibaraki 305-0801, Japan}\\
  {\em E-mail: takeshi.komatsubara@kek.jp}
  }
\maketitle
\baselineskip=11.6pt

\begin{abstract}
Recent results and future prospects 
in the rare kaon-decay experiments at KEK, BNL, CERN and FNAL
are reviewed. 
Emphasis is placed on the quark flavor dynamics 
in the flavor-changing neutral current processes.
Experimental and data-analysis techniques
developed in the rare decay experiments are discussed. 

\end{abstract}
\baselineskip=14pt

\section{Introduction}
The history of rare kaon-decay physics began in 1964 
with the unexpected evidence for the CP violating 
$K^0_L\to\pi^+\pi^-$ decay\cite{fitch} at the sensitivity of $10^{-3}$. 
This discovery, which in later stimulated Kobayashi and Maskawa\cite{KM} 
to introduce the third generation of quarks and leptons, 
was one of the major contributions 
of kaon physics toward  establishing the Standard Model (SM). 
Another famous contribution was the absence of kaon decays due to
flavor-changing neutral current (FCNC),  
$K^0_L\to\mu^+\mu^-$, $K^+\to\pi^+ e^+ e^-$ and $K^{+}\to\pi^+\nu\bar{\nu}$, 
in 1960's.
The GIM mechanism\cite{GIM} explained it with the unitary matrix 
for weak-eigenstate mixing. 
All of these had been achieved
in advance of the charm-quark discovery 
in 1974. 

The next theoretical milestone emphasized here was 
so-called ``Inami-Lim loop functions''\cite{IL} in 1981 
for FCNC processes with heavy quarks and leptons.
In addition to the continual searches for decays  
at higher level than the SM predictions,
the anticipation of detecting rare kaon decays in the SM range 
got more and more realistic with the rise of the top-quark mass.
The first evidence for the $K^+\to\pi^+\nu\bar{\nu}$ decay
reported by the E787 collaboration\cite{pnn1-1} in 1997
opened a new era of ``testing the SM by measuring rare processes''.
In the current experiments (tab.\ref{kaondecay}), 
with millions of kaon decays per second 
by high-intensity proton synchrotrons,
searches and measurements 
with the sensitivity of $10^{-7}\sim 10^{-12}$ are being performed; 
these are the frontiers that no other heavy-flavor physics can reach at present.
\begin{table}[ht]
  \centering
  \caption{ \it Rare kaon-decay experiments being reviewed in this article.
    ``${\surd}$'' means  data taking of the experiment is completed.
    ``${\ast}$'' means  detector construction of the experiment is not started.
    }
  \vskip 0.1 in
  \begin{tabular}{|l|lr|l|l|}
 \hline
 lab & \multicolumn{2}{|l|}{accelerator} & experiment & kaon decay\\
 \hline\hline
 KEK & PS  		&(12 GeV)	&E246 $^{\surd}$			& stopped ${K^+}$\\
     &                 	&		&E391a				& ${K^0_{L}}$\\
 BNL & AGS 		&(25 GeV)	&E787 $^{\surd}$\ /\ E949	& stopped ${K^+}$\\
     &          	&		&E865 $^{\surd}$  		& in-flight ${K^+}$\\
     &    		&		&E871 $^{\surd}$  		& ${K^0_{L}}$\\
     &    		&		&KOPIO $^{\ast}$			& ${K^0_{L}}$\\
 CERN& SPS		&(450 GeV)	&NA48    			& ${K^0_{L}}$, ${K^0_{S}}$\\
 FNAL& Tevatron		&(800 GeV)	&KTEV $^{\surd}$    		& ${K^0_{L}}$\\
     & Main Injector	&(120 GeV)	&CKM $^{\ast}$			& in-flight ${K^+}$\\
 \hline
  \end{tabular}
  \label{kaondecay}
\end{table}

In this article, 
recent results and future prospects 
in the rare kaon-decay experiments are briefly reviewed. 
``$\epsilon^{\prime}/\epsilon$'' CP violation in kaon decays
is reviewed by E.~Cheu\cite{Cheu}.
For further and latest information on the field,
in particular on the radiative decays and chiral dynamics
which are not covered here, 
the article by A.R.~Barker and S.H.~Kettell\cite{BarkerKettell}
and the Web site of the KAON2001 conference\cite{Kaon2001} are recommended. 

Remind that the experimental upper limits in this article are 
at 90\% confidence level.

\section{FCNC}
\begin{figure}[ht]
  \vspace{7.0cm}
  \includegraphics{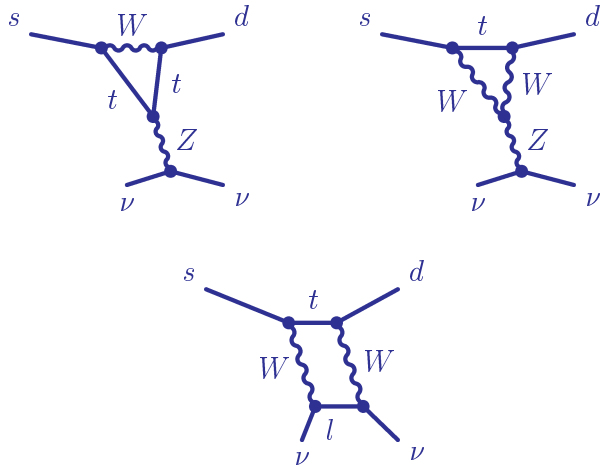}
  \caption{\it
    Penguin and Box diagrams in $K\to\pi\nu\bar{\nu}$.
    \label{Penguin} }
\end{figure}
The FCNC process in kaon decays is
strange-quark to down-quark transition and
is induced in the SM by the $W$ and $Z$ loop effects 
as Penguin and Box diagrams (fig.\ref{Penguin}). 
The top-quark in the loops dominates the transition 
because of its heavy mass, and
the quantity $\lambda_{t}$: 
\begin{equation}
  \lambda_{t} \equiv V_{ts}^{*} \cdot V_{td} =  
      A^{2} \lambda^5 \cdot (1 - \rho - i \eta),
\end{equation}
where $A$, $\lambda$, $\rho$ and $\eta$ are 
the Wolfenstein parameterization of the 
Cabibbo-Kobayashi-Maskawa(CKM) matrix,
is measured.  
The decays are rare due to $\lambda^5$, and 
are precious because the important parameters $\rho$ and $\eta$
can be determined from them. 
The decay amplitude of $K^0_L$ is 
a superposition of the amplitudes of $K^0$ and $\bar{K}^0$
and is proportional to the imaginary part of $V_{td}$ 
(and to $\eta$); 
observation of rare $K^0_L$ decays, in particular 
the $K^0_L\to\pi^0\nu\bar{\nu}$ decay,
is a new evidence for direct CP violation. 

Theoretical calculations are simple
if the kaon decay accompanies neutrinos \footnote{
 Long-distance contributions are negligible, and the hadronic matrix element 
 is extracted from the $K\to\pi e \nu$ decay.}. 
For charged leptons in the final state, 
the transition is also induced 
by long-distance effects with $\gamma$ emission 
due to hadronic interactions. 
Rare kaon decays with charged leptons 
are easier to detect in experiments 
but have difficulties in theoretical interpretation. 
For example,
a precise measurement of the branching ratio
of the $K^0_L \to \mu^+\mu^-$ decay
$(7.18\pm0.17)\times 10^{-9}$ was achieved 
by the E871 collaboration\cite{E871mumu} 
based on the 6.2K signal events.
However, 
the decay mode is saturated by an absorptive process: 
$K^0_L\to\gamma\gamma$ and the two $\gamma$'s 
subsequently scattered into muons. 
The estimated branching ratio due to this process, 
called ``unitarity bound''\cite{Sehgal},
is $(7.07\pm0.18)\times 10^{-9}$.
In this decay mode we basically look at a QED process
and cannot get good information on the CKM matrix elements. 

An idea is to use $K^0_S$ decays; 
when a CP violating effect is observed in $K^0_L$, 
the effect can be cross-checked by a null result 
in the corresponding $K^0_S$ decay.
An example is a large CP-violating asymmetry 
$(13.6\pm 2.5_{stat}\pm 1.2_{syst})$\%
observed by the KTEV collaboration\cite{KTEVppee}
in the distribution of 
$K^0_L\to\pi^+\pi^- e^+ e^-$ decays,
whose branching ratio was $(3.2\pm 0.6_{stat} \pm 0.4_{syst})\times 10^{-7}$, 
in the angle between the decay planes of the $e^+ e^-$ and $\pi^+\pi^-$ pairs
in the $K^0_L$ rest frame.
The asymmetry was confirmed by a preliminary result 
from the NA48 collaboration\cite{NA48ppee},
$(13.9\pm 2.7_{stat}\pm 2.0_{syst})$\%.
NA48 also measured the asymmetry in the distribution of 
$K^0_S\to\pi^+\pi^- e^+ e^-$ decays,
whose branching ratio was $(4.3\pm 0.2_{stat} \pm 0.3_{syst})\times 10^{-5}$,
with the same detector.
No asymmetry $(-0.2\pm 3.4_{stat}\pm 1.4_{syst})$\% was observed, 
which indicates that 
the large asymmetry in $K^0_L$ is not an effect due to final-state interactions
nor an artifact due to acceptance errors in the measurement.

\section{$K^0_L\to\pi^0 e^+ e^-$}
\begin{figure}[ht]
  \vspace{7.0cm}
  \includegraphics{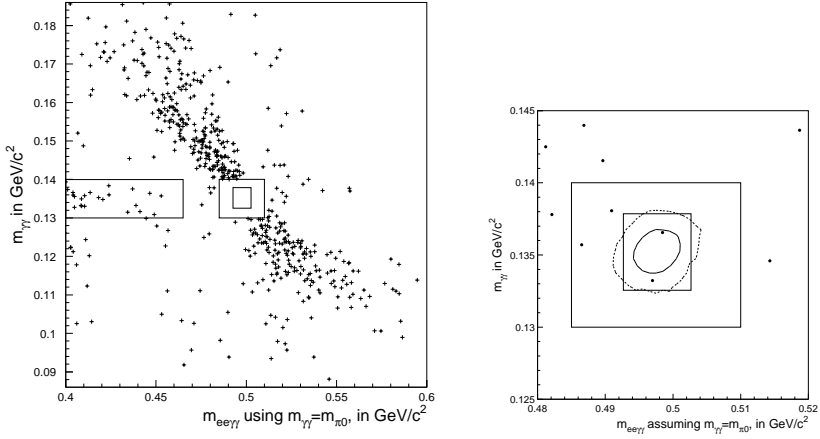}
  \caption{\it
    KTEV result on $K^0_L\to\pi^0 e^+ e^-$: 
    the distribution of 
    reconstructed two photon mass $m_{\gamma\gamma}$ vs
    reconstructed mass of the four particle system $m_{ee\gamma\gamma}$ 
    (left)
    and the same distribution on an expanded scale around the signal region
    after phase space fiducial cuts
    were imposed (right). 
    \label{pieeplot} }
\end{figure}

An upper limit on the branching ratio 
of the $K^0_L \to \pi^0 e^+e^-$ decay 
$< 5.1 \times 10^{-10}$ was set 
by the KTEV collaboration\cite{KTEVpiee} 
with their 1997 data set \footnote{
  KTEV has the 1999 data set, which is being analyzed.} 
based on two events in the signal region 
with the background estimate of $1.06\pm 0.41$ events
(fig.\ref{pieeplot}). 
The limiting background was 
the radiative Dalitz decay 
$K^0_L\to e^+ e^- \gamma \gamma$, 
whose branching ratio was $(6.9\pm 1.0)\times 10^{-7}$\cite{PDG},
when $m_{\gamma\gamma}$ was equal to the $\pi^0$ mass.
In the analysis, the ``blind'' region
($m_{\gamma\gamma}=135\pm5$ MeV/$c^2$, $485<M_{ee\gamma\gamma}<510$ MeV/$c^2$)
and the signal region 
($m_{\gamma\gamma}=135.20\pm2.65$ MeV/$c^2$, 
 $m_{ee\gamma\gamma}=497.67.20\pm5.00$ MeV/$c^2$)
were blanked out until the analysis procedure 
and selection criteria (``cuts'') were finalized.

\begin{figure}[ht]
  \vspace{6.0cm}
  \includegraphics{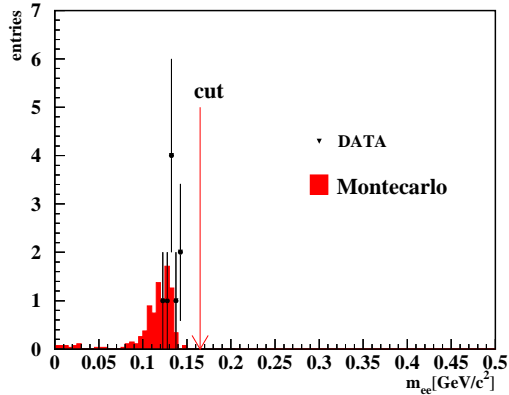}
  \caption{\it
    NA48 result on $K^0_S\to\pi^0 e^+ e^-$: 
    reconstructed invariant mass distribution of the
    $e^+e^-$ pair selected as $K^0_S\to\pi^0 e^+ e^-$. 
    The cut was imposed to remove the background from 
    the $\pi^0\to e^+ e^- \gamma$ decay.
    \label{NA48plot} }
\end{figure}

There are three contributions to the $K^0_L \to \pi^0 e^+e^-$ decay: 
a direct CP-violating contribution,
a CP conserving contribution through $\pi^0\gamma^{\ast}\gamma^{\ast}$ 
intermediate states 
and an indirectly CP-violating contribution 
due to the $K_{1}$ component of $K^0_{L}$.
The first contribution was predicted to be 
$(4.3\pm 2.1)\times 10^{-12}$\cite{Buras} in the SM
and the second contribution was estimated to be 
around $5\times 10^{-12}$\cite{Valencia}
from measurements of the 
$K^0_L \to \pi^0 \gamma\gamma$ decay\cite{KTEVpigg}\cite{NA48ppee}.
The CP conserving $K^0_S\to \pi^0 e^+ e^-$ decay helps  
to do reliable estimation of the third contribution \footnote{
 The branching ratio for $K^0_L$ is 0.3\% of the branching ratio
 for $K^0_S$ because of $|\epsilon|^2$ and the ratio of their lifetimes.
}.
An upper limit on the branching ratio 
of the $K^0_S \to \pi^0 e^+e^-$ decay
$< 1.4 \times 10^{-7}$ was obtained 
by the NA48 collaboration\cite{NA48pieepub},
using data collected in 1999 during a 40-hour run
with a high-intensity $K^0_S$ beam,
based on zero events in the signal region 
with the background estimate of $<0.15$ events
(fig.\ref{NA48plot}). 
In the SM the branching ratio of the $K^0_S\to \pi^0 e^+ e^-$ decay
is predicted to be $10^{-9}\sim 10^{-8}$\cite{DAmbrosio}.
NA48 is going to take $K_{S}$ data in 2002 
with higher intensity and new target station, 
and get $\times 50$ improvement
on the decay mode. 

\section{$K^+\to\pi^+\nu\bar{\nu}$}
The branching ratio of the $K^+\to\pi^+\nu\bar{\nu}$ decay
is represented in the SM as: 
\begin{equation}
 B(K^+\to\pi^+\nu\bar{\nu}) = \\
 4.11 \times 10^{-11} \times A^{4}\times X(x_{t})^{2}
 \times [\ (\rho_{0}- \rho )^{2}\ +\ \eta^{2}\ ]
\end{equation}
where $X(x_{t})$ is the Inami-Lim loop function
with the QCD correction,
$x_t \equiv m_t^2/m_W^2$ and
$\rho_0$ is estimated to be $1.4\sim 1.6$ \footnote{
 $\rho_0-1$, without which the branching ratio
 should be proportional to $|V_{td}|^2$, 
 is due to the charm-quark contribution.}.
The theoretical uncertainty is 7\%
from the charm-quark contribution in the next-to-leading-logarithmic
QCD calculations\cite{BBL96}.
With the $\rho$-$\eta$ constraints from other kaon and B-meson decay
experiments, the SM prediction of the branching ratio is
$(0.75\pm 0.29)\times10^{-10}$\cite{Buras}.
Using only the results
on $B_d-\bar{B_d}$ and $B_s-\bar{B_s}$ mixing,
a branching ratio limit $<1.15\times 10^{-10}$ can be extracted\cite{BB99}.
New physics beyond the SM could affect the branching ratio\cite{beyondSM}.
In addition, the two-body decay $K^+\to\pi^+X^0$,
where the $X^0$ is a weakly-interacting light particle
such as a familon\cite{familon},
could also be observed as a ``$\pi^+$ plus nothing'' decay
with a monochromatic pion.
Since the effects of new physics are not expected to be too large,
a precise measurement of a decay at the level of $10^{-10}$ is required.

\begin{figure}[ht]
% \vspace{7.0cm}
  \vspace{6.5cm}
  \includegraphics{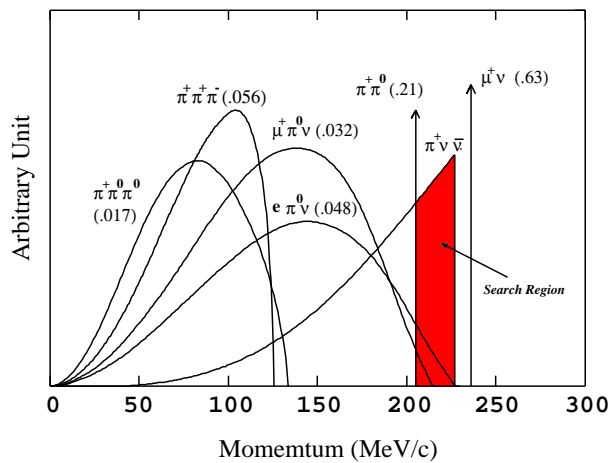}
  \caption{\it
    Momentum spectrum of the charged particles from $K^+$ decays at rest.
    \label{spctl} }
\end{figure}
The E787 and E949 collaborations for a study of 
$K^+\to\pi^+\nu\bar{\nu}$ and related decays 
measure the charged track emanating from $K^+$ decays at rest.
The $\pi^+$ momentum from $K^+\to\pi^+\nu\bar{\nu}$
is less than 227MeV/$c$, 
while the major background sources of
$K^+\to\pi^+\pi^0$ ($K_{\pi 2}$, 21.2\%) and
$K^+\to\mu^+\nu$   ($K_{\mu 2}$, 63.5\%)
are two-body decays and have monochromatic momentum of
205MeV/$c$ and 236MeV/$c$, respectively
(fig.\ref{spctl}). 
The region ``above the $K_{\pi 2}$'' between 211MeV/$c$ and 229MeV/$c$
is adopted for the search.
Background rejection is essential in this experiment,
and the weapons for
redundant kinematics measurement,
$\mu^+$ rejection, and extra-particle and photon veto
are employed.
Each weapon should have a rejection of $10^5\sim 10^6$,
and reliable estimation of these rejections using real data is 
the key of the experiment.

\begin{figure}[ht]
  \vspace{9.0cm}
  \includegraphics{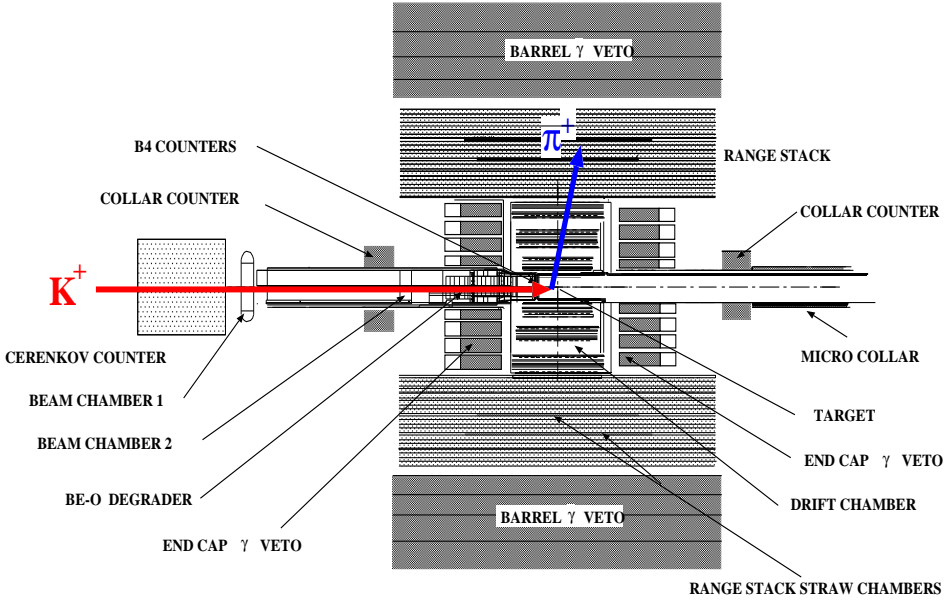}
  \caption{\it
    Schematic side-view of the E787/E949 detector at BNL.
    \label{detector} }
\end{figure}
The E787/E949 detector (fig.\ref{detector}) 
is a solenoidal spectrometer
with the 1.0-Tesla field directed along the beam line.
Slowed by a BeO degrader, kaons of about 700MeV/$c$ from AGS
reach the scintillating-fiber target
at the center of the detector and decay at rest.
A delayed coincidence requirement ($>$ 2nsec) of the timing
between the stopping kaon and the outgoing pion
helps to reject backgrounds of pions scattered into the detector
or kaons decaying in flight.
Charged decay products pass through the drift chamber,
lose energy by ionization loss and come to rest
in the Range Stack made of plastic scintillators and straw chambers.
Momentum, kinetic energy and range are measured
to reject the backgrounds by kinematic requirements.
For further rejection of $\mu^+$ tracks from $K_{\mu 2}$
the decay chain $\pi^+\to\mu^+\to e^+$ is identified 
in the Range Stack counter in which the $\pi^+$ comes to rest,
using output pulse-shape information of the counter.
$K_{\pi 2}$ and other decay modes with extra particles
(photon, $e$, ...) are vetoed by the in-time signals
in the hermetic shower counters.

\begin{figure}[ht]
  \vspace{7.5cm}
  \includegraphics{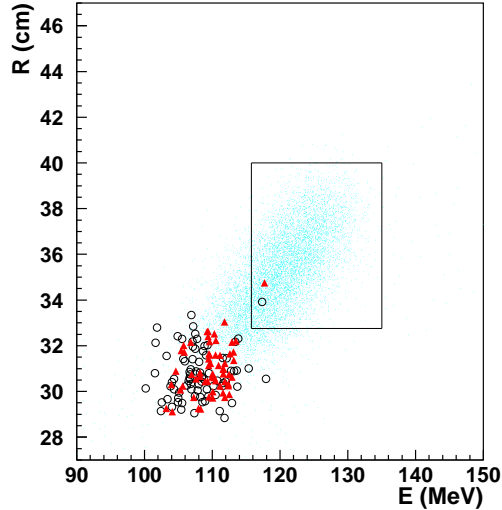}
  \caption{\it
    E787 result on $K^+\to\pi^+\nu\bar{\nu}$: 
    range vs kinetic energy plot of the final sample.
    The circles are for the 1998 data and the triangles are
    for the 1995-1997 data set. The simulated distribution
    of expected events from $K^+\to\pi^+\nu\bar{\nu}$ is indicated 
    by dots. 
    The box indicates
         the signal acceptance region.
    \label{pinnplotnew} }
\end{figure}
Final results from E787 were announced\cite{pnn1-final} 
about five months after the XXI Physics in Collision conference
was held, 
including a new 1998 data set of comparable sensitivity to   
the 1995-1997 data set already reported\cite{pnn1-2}.
One new event was observed in the new data set,
bringing the total for the combined data set to two
(fig.\ref{pinnplotnew}).
Including all data taken,
the backgrounds were estimated to contribute $0.15\pm 0.05$ events.
The branching ratio of the $K^+\to\pi^+\nu\bar{\nu}$ decay was 
$1.57^{+1.75}_{-0.82}\times 10^{-10}$.
A constraint $2.9\times 10^{-4} < |\lambda_{t}| < 1.2\times 10^{-3}$
was provided without reference to the B-meson system. 

 The E949 experiment\cite{E949} 
 continues the study at the AGS based on the experience of E787.
 E949 is expected to reach a sensitivity of $(8-14)\times 10^{-12}$
 in $\sim$2 years of running and determine $|\lambda_{t}|$ to $20-30$\%.
 Detector upgrade (additional shower counters in the barrel and beam regions
  to improve photon detection, 
  new beam counters, 
  new trigger system, Range Stack readout by TDC for deadtime reduction, ...)
 is completed,
 and the physics run starts in 2001.

\section{$K^0_L\to\pi^0\nu\bar{\nu}$}
The branching ratio of the neutral counterpart, 
the $K^0_L\to\pi^0\nu\bar{\nu}$ decay,
is represented in the SM as: 
\begin{equation}
 B(K^0_L\to\pi^0\nu\bar{\nu}) = \\
 1.80 \times 10^{-10} \times A^{4}\times X(x_{t})^{2}
 \times \eta^{2}
\end{equation}
and the SM prediction is
$(2.6\pm 1.2)\times10^{-11}$\cite{Buras}.
A model-independent bound\cite{GrossmanNir}
\begin{equation}
 B(K^0_L\to\pi^0\nu\bar{\nu}) < 
  4.4 \times B(K^+\to\pi^+\nu\bar{\nu})
   < 2.6 \times 10^{-9}
\end{equation}
can be extracted 
from its isospin-relation to the $K^+\to\pi^+\nu\bar{\nu}$ decay.

In comparison with the $K^+$ case, 
$K^0_L\to\pi^0\nu\bar{\nu}$ has some advantages. 
 \begin{itemize}
  \item The theoretical uncertainty, $\sim$1\%, is smaller. 
  \item No $K_{\mu 2}$($K^0_L\to\mu^+\mu^-$) background has to be worry about. 
  \item $K_{\pi 2}$($K^0_L\to\pi^0\pi^0$) has been suppressed to $10^{-3}$.
 \end{itemize}
However, the following severe difficulties exists.
 \begin{itemize}
  \item The signal is an in-flight $K^0_L$ decay into ``$\pi^0$ plus nothing''
        in a neutral beam line with plenty amount of $\pi^0$'s coming 
         from $K^0_L$ decays and neutrons, which could easily create 
         $\pi^0$'s with residual gas. 
  \item No kinematic constraint (initial kaon, decay vertex, ...) is available 
        with the usual techniques for neutral kaon experiments. 
  \item The expected branching ratio is low enough.
 \end{itemize}
To beat the major background from $K^0_L\to\pi^0\pi^0$
in case two out of four photons are missed, 
photon detection with quite-low inefficiency ($< 10^{-3}\sim 10^{-4}$) is
required to the detector. 

\begin{figure}[ht]
  \vspace{7.0cm}
  \includegraphics{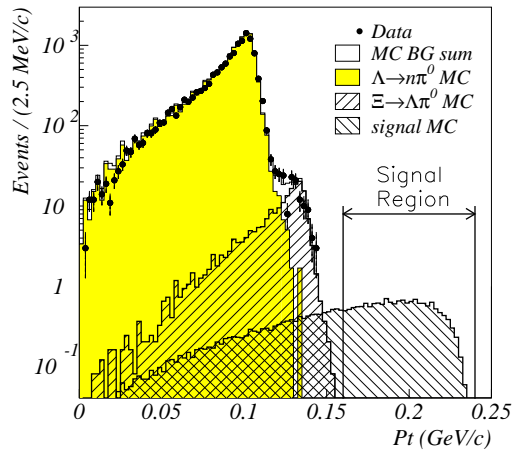}
  \caption{\it
    KTEV result on $K^0_L\to\pi^0 \nu\bar{\nu}$: 
    the distribution of 
    total momentum transverse to the $K^0_L$ flight direction.
    \label{pinnplot} }
\end{figure}
The current best upper limit on the branching ratio 
of the $K^0_L \to \pi^0\nu\bar{\nu}$ decay 
$< 5.9 \times 10^{-7}$ was set 
by the KTEV collaboration\cite{KTEVpinndalitz} 
with their 1997 data set
based on zero events in the signal region 
with the background estimate of $0.12^{+0.05}_{-0.04}$ events
(fig.\ref{pinnplot}). 
The Dalitz decay mode $\pi^0\to e^+e^-\gamma$ (1.2\%)
for the final state of $K^0_L\to\pi^0\nu\bar{\nu}$ 
was used in this search.
However, to reach the SM sensitivity, 
reconstructing $\pi^0\to \gamma\gamma$ from the $K^0_L\to\pi^0\nu\bar{\nu}$ 
has to be considered.
A search for $K^0_L\to\pi^0\nu\bar{\nu}$ with $\pi^0\to \gamma\gamma$
was performed by the KTEV collaboration\cite{KTEVpinngg} 
with their one-day special run; 
an upper limit on the branching ratio was determined to be 
$< 1.6 \times 10^{-6}$
 based on one event in the signal region 
 with the background estimate of $3.5\pm 0.9$ events.

\section{Future prospects in $K\to\pi\nu\bar{\nu}$ experiments}

Measurement of $B(K^0_L\to\pi^0\nu\bar{\nu})$
and $B(K^+\to\pi^+\nu\bar{\nu})$  
is the issue of kaon physics 
in the next five to ten years. 
In the $\rho$-$\eta$ plane, 
the height of the ``unitarity triangle''
is proportional to 
$B(K^0_L\to\pi^0\nu\bar{\nu})^{1/2}$ 
and a side of it is 
to $B(K^+\to\pi^+\nu\bar{\nu})^{1/2}$.
The area of the triangle is proportional to
so-called ``Jarlskog invariant''\cite{Jarlskog}
in kaon sector: 
\begin{equation}
 J_{CP}=Im(V_{ts}^{\ast}\cdot V_{td}
     \cdot V_{us} \cdot V_{ud}^{\ast})
                 =\lambda (1-\frac{\lambda^2}{2})\times
                  Im(\lambda_{t})
\end{equation}
and is directly related to these decay modes \footnote{
  A constraint on $|\lambda_{t}|$ from
  $K^+\to\pi^+\nu\bar{\nu}$ sets an upper limit on
  $Im(\lambda_{t})$.
  }. 
If both branching ratios are measured  
with $\sim$10\% precision ($\sim$100 signal events)
with highly sophisticated and special-purpose detectors, 
the triangle is determined with good precision 
only from the information in kaon sector. 
Comparing with the triangle by B-meson system, 
it can be tested 
whether the source of CP violation is only from the CKM phase or not.

\begin{figure}[ht]
  \vspace{7.0cm}
  \includegraphics{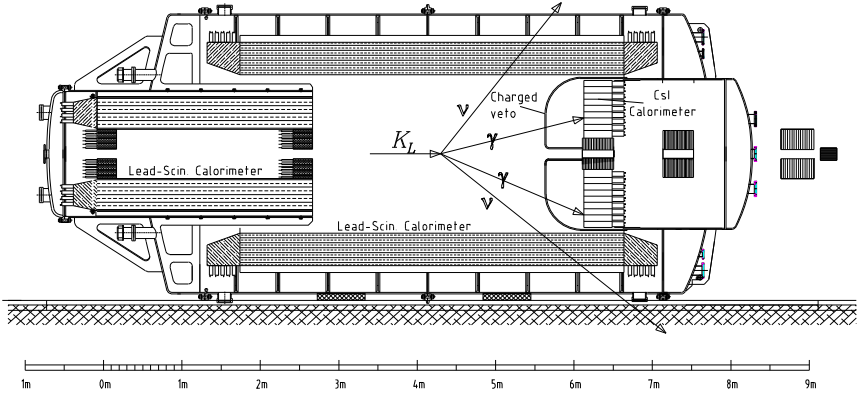}
  \caption{\it
    Side-view of the E391a detector at KEK under construction.
    \label{e391adet} }
\end{figure}
The E391a experiment\cite{E391a} (fig.\ref{e391adet})
is the first dedicated search
for the $K^0_L\to\pi^0\nu\bar{\nu}$ decay.
A collimated ``pencil'' neutral beam is designed carefully.
A calorimeter with CsI crystals detects two photons from $\pi^0$
and measures their energy and position.
The $K^0_L$-decay vertex position along the beam line 
is determined from the constraint of $\pi^0$ mass. 
Calorimeters which cover the decay region intend to do
hermetic photon veto and reject the background from $K^0_L\to\pi^0\pi^0$.
Beam line survey and detector construction were started,
and the data taking is scheduled from 2003.
The goal of E391a is to reach a sensitivity at $10^{-10}$,
and they plan to continue the study at the new 50GeV
proton synchrotron in the 
High Intensity Proton Accelerator Facility\cite{JHF},
which started construction in Japan this year. 

\begin{figure}[ht]
  \vspace{6.0cm}
  \includegraphics{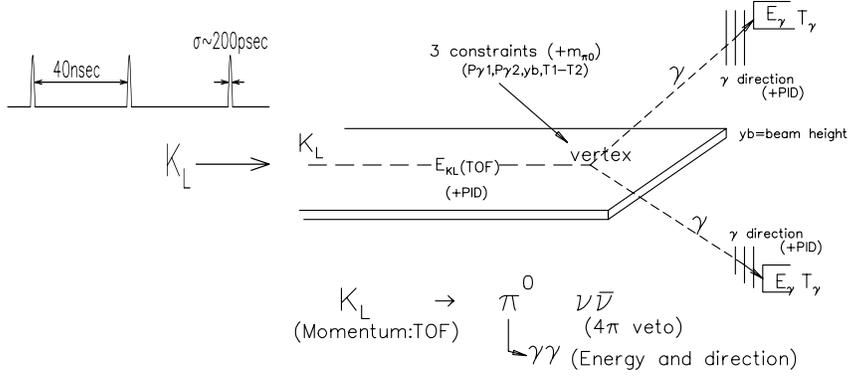}
  \caption{\it
    Principles of the proposed KOPIO detector at BNL. 
    \label{kopiodet} }
\end{figure}
More ambitious proposal is the KOPIO experiment\cite{KOPIO}
(fig.\ref{kopiodet}),
whose principles are to give kinematic constraints to the $K_L$ decay 
as much as possible. 
A RF-bunched proton beam from AGS makes low-energy $K^0_L$'s
(around 800MeV/$c$) with a large targeting angle, 
so that the momentum of each kaon is measured 
with TOF technique and the decay can be analyzed 
in the $K^0_L$ rest frame.
A combination of the pre-radiator and Shashlik calorimeter
intends to measure the timing, energy, position and angle
of low energy photons and fully reconstruct the decay. 
The goal of KOPIO is to observe 50 signal events in the SM
with a signal-to-background ratio of 2. 

\begin{figure}[ht]
  \vspace{8.5cm}
  \includegraphics{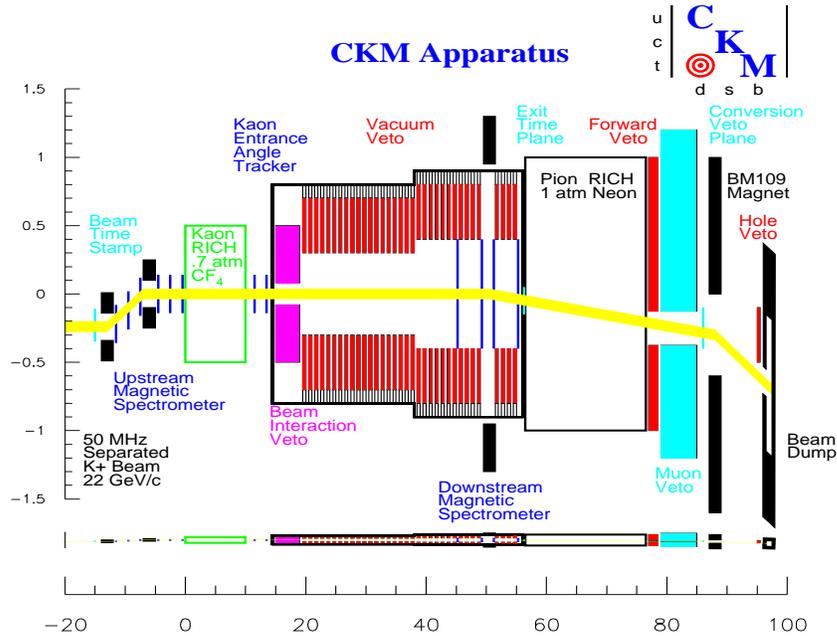}
  \caption{\it
    Plan-view of the proposed CKM detector at FNAL.
    The lower section shows the true proportions of the apparatus.
    \label{ckmdet} }
\end{figure}
For $K^+\to\pi^+\nu\bar{\nu}$, 
the CKM experiment\cite{CKM}
(fig.\ref{ckmdet})
intends to study the decay mode 
with in-flight $K^+$ decays for the first time.  
A RF-separated 22GeV/$c$ $K^+$ beam 
and RICH technique for particle identification 
are used. 
The goal of CKM is to collect 100 signal events in the SM
with 10 background events.

\section{Blind analysis}

A technique of ``blind analysis'' is widely used 
in the field of rare kaon-decay experiments.
The principal idea is to hide the answer in order to avoid human bias. 
The procedure is as follows. 
At first, the signal region (``BOX'') in multi-dimensional
``cut'' space \footnote{
 Not only the kinematic variables in the final plot 
 but also some important variables
 for timing, visible energy and so on can be used 
 to define the signal region.} 
is pre-determined.
Other cuts are chosen without looking at the events
inside the BOX, and are frozen. 
Background estimates in the regions ``outside-the-BOX'',
where the background-to-signal ratio
is expected to be higher than that in the signal region, 
are compared with the number of events observed in them, 
and in the final step the BOX was opened and the number of signal events
is counted.
With the same philosophy, 
in measuring a branching ratio, asymmetry, polarization 
and so on, 
some blind numbers are added or multiplied until looking at the result
in the final step.
 
One of the main reasons to take blind analysis
in rare decay searches is, 
we expect only a handful signal events, even if we succeed
to detect, 
and we should be careful to avoid playing with them
(adjust a cut to remove or save a candidate event, ...)
in the analysis.

\section{T-violating transverse muon polarization in $K^+\to\pi^0\mu^+\nu$}

Though the $K^+\to\pi^0\mu^+\nu$ itself 
(branching ratio $=$ $3.18\pm0.18$\%\cite{PDG})
is not rare, 
here is a good example of looking beyond the SM
by precisely measuring a decay property with high statistics. 
The transverse muon polarization $P_T$ of the $K^+\to\pi^0\mu^+\nu$ decay
(the perpendicular component of muon spin vector relative to the decay plane
 determined by the momentum vectors of muon and pion 
 in the $K^+$ rest frame) is a T-odd quantity,
 and a nonzero value of $P_T$ indicates
 violation of T invariance.
 Any spurious effect from final-state interactions is small,
 because no charged particle other than muon exists in the final state.
 $P_T$ due to the CP violation in the SM is as small as $10^{-7}$\cite{Sanda},
 while new sources of CP violation could appear in the polarization 
 at the level of $10^{-3}$.

The E246 collaboration\cite{E246pub} measures the charged track and photons
from $K^+$ decays at rest
with the superconducting toroidal spectrometer 
(consisting of 12 identical, 0.9-Tesla spectrometers arranged in 
 rotational symmetry),
which enables to control possible sources of the systematic error 
in polarization measurement. 
No transverse polarization 
$(-0.33\pm 0.37_{stat}\pm 0.09_{syst})$\% was observed
by a preliminary result
from the the E246 collaboration\cite{Imazato}
with their 1996-1998 data set.
Combined with the 1999-200 data set, which is being analyzed, 
the error in $P_T$ is expected to be reduced to $0.003$.

 E246 reported the form factors 
 of the $K^+\to \pi^0 e^+ \nu$ decay\cite{E246form}.
 The PDG's combined results\cite{PDG}
 for the ratio of the strengths of scalar and tensor couplings 
 to that of the vector coupling
 ($f_{S}/f_{+}(0)$ and $f_{T}/f_{+}(0)$, respectively)
 differ from zero and contradict the V-A interaction in the SM. 
 After analyzing the observed Dalitz plots containing 41 K events
 from their special trigger runs in 1996-1997
 with two sets of magnetic field (0.65 Tesla and 0.9 Tesla) 
in the spectrometer, 
 $f_S/f_+(0)=-0.002\pm 0.026_{stat}\pm 0.014_{syst}$
 and $f_T/f_+(0)=-0.01\pm 0.14_{stat}\pm 0.09_{syst}$
 were determined.
 These values are consistent with the SM prediction of zero. 
 E246 also reported the ratio of the 
 $K^+\to\pi^0\mu^+\nu$ and $K^+\to \pi^0 e^+ \nu$ 
 decay widths\cite{E246ratio}
 $0.671\pm 0.007_{stat}\pm 0.008_{syst}$.

\section{Lepton flavor violation in kaon decays}

Lepton flavor (LF) violation has been studied for more than 50 years,
and is a current topic of particle physics 
because of recent results on neutrino oscillation and 
theoretical predictions to the processes
induced by Supersymmetric loop effects.
There are new proposals 
to search for the $\mu^+\to e^+\gamma$ decay at $10^{-14}$ in PSI\cite{PSI} 
and the $\mu^{-}N\to e^{-}N$ conversion 
at $5\times 10^{-17}$ in BNL\cite{MECO}.

Experimental search for LF violation in kaon decays 
($K^0_L\to\mu^{\pm} e^{\mp}$, $K^0_L\to\pi^0 \mu^{\pm} e^{\mp}$, 
 $K^+\to\pi^+\mu^+ e^-$) \footnote{
  Both two-body and three-body decays have to be explored 
  in spite of the phase-space difference, 
   because the $K\to\pi\mu e$ decay is sensitive 
   to a vector or scaler current.}
also has a long history.
The experimental limit on neutrino masses and mixing 
more than 10 years ago, unfortunately, had already restricted 
the branching ratio of the $K_L\to\mu^{\pm} e^{\mp}$ decay,
for example, to be less than $10^{-15}$
even with possible Left-Right asymmetry\cite{Langacker}.
Introducing Supersymmetric models would not change the situation
so much because ``Super-GIM'' suppression mechanism is 
expected for both quark and lepton sectors\cite{SUSY}.
The advantage of kaon decays is to be able to investigate
the LF violating process involving both 
quarks and leptons with the highest sensitivity.
Assuming a number for quarks and leptons in the same generation
(1 for down-quark and electron, 2 for strange-quark and muon, ...)
due to Horizontal Symmetry or Compositeness or Leptoquarks, 
the net number is conserved in the LF-violating kaon decays. 
The mass of hypothetical generation-changing
gauge boson for tree-level effects should 
be a few hundred TeV scale\cite{CHS}. 

\begin{figure}[ht]
  \vspace{5.0cm}
  \includegraphics{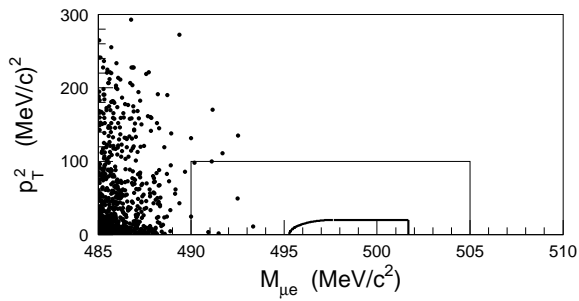}
  \caption{\it
     E871 result on $K^0_L\to\mu^{\pm} e^{\mp}$: 
     plot of the two-body transverse momentum squared $p_{T}^2$
      versus the two-body invariant mass $M_{\mu e}$.
     The exclusive region for the blind analysis is indicated by the box. 
     The signal region is indicated by the smaller contour.
    \label{E871mueplot} }
\end{figure}
The best upper limit on the branching ratio 
of the $K^0_L\to\mu^{\pm} e^{\mp}$ decay
$< 4.7 \times 10^{-12}$ \footnote{
   This is the most stringent upper limit to particle decays.}
was set 
by the E871 collaboration\cite{E871mue} 
based on zero events in the signal region 
with the background estimate of $0.1$ events
(fig.\ref{E871mueplot}).  
An upper limit on the branching ratio 
of the $K^0_L\to\pi^0 \mu^{\pm} e^{\mp}$ decay
$< 4.4 \times 10^{-10}$ was set 
by a preliminary result
from the KTEV collaboration\cite{KTEVpimue}
with their 1997 data set
based on two events in the signal region 
with the background estimate of $0.61\pm 0.56$ events.
An upper limit on the branching ratio 
of the  $K^+\to\pi^+\mu^+ e^-$ decay
$< 2.8 \times 10^{-11}$ was set 
by the E865 collaboration\cite{E865pimue} 
with their 1995-1996 data set
(zero events in the signal region)
and the result of an earlier experiment E777\cite{E777}.
Combined with the 1998 data set, which is being analyzed, 
the limit is expected to reach $\sim 1.5\times 10^{-11}$.

 E865 reported the upper limit on the branching ratio
of the $K^+\to\mu^+\mu^+\pi^-$ decay\cite{E865db}
$< 3.0 \times 10^{-9}$.
This decay is a neutrino-less ``double muon'' decay 
changing total lepton number by two and 
provides a unique channel to search for effects of 
Majorana neutrinos in the second generation.

\section{Summary}
\begin{table}[ht]
  \centering
  \caption{ \it Progress in the field of kaon rare decays.
    }
  \vskip 0.1 in
 \begin{tabular}{|ll|r|r|r|}
  \hline
    decay modes                       &                 &PDG-86           &PDG-96           & PIC-01 \\
  \hline\hline
   $K^0_L\to\mu^+\mu^-$               &in $10^{-9}$     &$9.1\pm 1.9$     &$7.2\pm 0.5$     &$7.18\pm 0.17$\\
  \hline
   $K^0_L\to\pi^0 e^+ e^-$            &in $10^{-10}$    &$<23000$         &$<43$            &$<5.1$\\
   $K^0_S\to\pi^0 e^+ e^-$            &in $10^{-7}$     &-                &$<11$            &$<1.4$\\
  \hline
   $K^+\to\pi^+\nu\bar{\nu}$          &in $10^{-10}$    &$<1400$          &$<24$            &$1.57^{+1.75}_{-0.82}$\\
   $K^0_L\to\pi^0\nu\bar{\nu}$        &in $10^{-7}$     &-                &$<580$           &$<5.9$\\
  \hline
   $K^0_L\to\mu^{\pm}e^{\mp}$         &in $10^{-12}$    &$<6000000$       &$<33$            &$<4.7$\\
   $K^0_L\to\pi^0\mu^{\pm}e^{\mp}$    &in $10^{-10}$    &-                &-                &$<4.4$\\
   $K^+\to\pi^+\mu^{+}e^{-}$          &in $10^{-11}$    &$<500$           &$<21$            &$<2.8$\\
  \hline
  \end{tabular}
  \label{summarytable}
\end{table}
Let us summarize the progress in the field of kaon rare decays 
in the last 15 years (tab.\ref{summarytable}). 
The experimental results reported in this article (PIC-01)
are compared to the results on Review of Particle Properties
in 1986 (PDG-86)\cite{PDG86}, 
when I started particle physics as a graduate student,
and in 1996 (PDG-96)\cite{PDG96},
when the first round of rare decay experiments are completed. 
The improvements were enormous between 1986 and 1996, 
and about an order of magnitude of improvements are achieved
in the last 5 years.
We do expect further improvements and potential discoveries 
by the next round of experiments
in the next five to ten years. 

\section*{Acknowledgments}
I would like to thank
 A.R.~Barker, D.A.~Bryman,
 A.~Ceccucci, E.~Cheu, P.S.~Cooper, 
 J.~Imazato, T.~Inagaki, 
 S.H.~Kettell, 
 G.Y.~Lim, L.S.~Littenberg, 
 W.R.~Molzon, 
 S.~Sugimoto, 
 T.~Yamanaka, 
  and
 M.E.~Zeller
for providing me help with my talk at the XXI Physics in Collision
conference and this article for the Proceedings. 
I would like to acknowledge support from 
Science and Technology Promotion Foundation of Ibaraki.
%

%
% References
%

%

\begin{thebibliography}{99}
\bibitem{fitch}J.H.~Christenson {\it et al}, 
  Phys. Rev. Lett. {\bf 13}, 138 (1964).
\bibitem{KM}M.~Kobayashi and T.~Maskawa,
  Progr. Theor. Phys. {\bf 49}, 652 (1973) 652.
\bibitem{GIM}S.L.~Glashow, J.~Iliopoulos and L.~Maiani,
  Phys. Rev. {\bf D2}, 1285 (1970); 
  M.M.~Gaillard and B.W.~Lee,
  Phys. Rev. {\bf D10}, 897 (1974).
\bibitem{IL}T.~Inami and C.S.~Lim,
  Progr. Theor. Phys. {\bf 65}, 297 (1981); 1172(E) (1981). 
\bibitem{pnn1-1}S.~Adler {\it et al}, 
  Phys. Rev. Lett. {\bf 79}, 2204 (1997).
\bibitem{Cheu}E.~Cheu, in these Proceedings.
\bibitem{BarkerKettell}A.R.~Barker and S.H.~Kettell,
  Annu. Rev. Nucl. Part. Sci. {\bf 50}, 249 (2000).
\bibitem{Kaon2001}KAON2001 International Conference on CP Violation
  (Pisa, June 2001),
  {\tt http://www.pi.infn.it/kaon2001/}.
\bibitem{E871mumu}D.~Ambrose {\it et al}, 
  Phys. Rev. Lett. {\bf 84}, 1389 (2000).
\bibitem{Sehgal}L.M.~Sehgal, 
  Phys. Rev. {\bf 183}, 1511 (1969).
\bibitem{KTEVppee}J.~Adams {\it et al}, 
  Phys. Rev. Lett. {\bf 80}, 4123 (1998); 
  A.~Alavi-Harati {\it et al}, 
  Phys. Rev. Lett. {\bf 84}, 408 (2000).
\bibitem{NA48ppee}M.~Martini,
  in KAON2001. 
\bibitem{KTEVpiee}A.~Alavi-Harati {\it et al}, 
  Phys. Rev. Lett. {\bf 86}, 397 (2001).
\bibitem{PDG}Particle Data Group, D.E.~Groom {\it et al}, 
  Eur. Phys. J. C {\bf 15}, 1 (2000).
\bibitem{Buras}A.J.~Buras, hep-ph/0101336 (2001); 
  A.J.~Buras and R.~Fleischer, hep-ph/0104238 (2001). 
\bibitem{Valencia}F.~Gabbiani and G.~Valencia, 
  Phys. Rev. D {\bf 64}, 094008 (2001).
\bibitem{KTEVpigg}A.~Alavi-Harati {\it et al}, 
  Phys. Rev. Lett. {\bf 83}, 917 (1999).
\bibitem{NA48pieepub}A.~Lai {\it et al}, 
  Phys. Lett. B {\bf 514} 253 (2001).
\bibitem{DAmbrosio}G.~D'Ambrosio {\it et al}, 
  JHEP {\bf 08}, 004 (1998).
\bibitem{BBL96}G.~Buchalla, A.J.~Buras and M.E.~Lautenbacher,
  Rev. Mod. Phys. {\bf 68} 1125 (1996).
\bibitem{BB99}G.~Buchalla and A.J.~Buras,
  Nucl. Phys. B {\bf 548} 309 (1999).
\bibitem{beyondSM}Y.~Nir and M.P.~Worah,
  Phys. Lett. B {\bf 423} 319 (1998)
  and references therein;
  A.J.~Buras {\it et al},
  Nucl. Phys. B {\bf 566} 3 (2000).
\bibitem{familon}F.~Wilczek,
  Phys. Rev. Lett. {\bf 49} 1549 (1982).
\bibitem{pnn1-final}S.~Adler {\it et al}, 
  hep-ex/0111091 (2001), to be published in Phys. Rev. Lett. 
\bibitem{pnn1-2}S.~Adler {\it et al}, 
  Phys. Rev. Lett. {\bf 84}, 3768 (2000).
\bibitem{E949}Brookhaven AGS Experiment E949: 
  {\tt http://www.phy.bnl.gov/e949/}.
\bibitem{GrossmanNir}Y.~Grossman and Y.~Nir,
  Phys. Lett. B {\bf 398} 163 (1997).
\bibitem{KTEVpinndalitz}A.~Alavi-Harati {\it et al}, 
  Phys. Rev. D {\bf 61}, 072006 (2000).
\bibitem{KTEVpinngg}J.~Adams {\it et al}, 
  Phys. Lett. B {\bf 447} 240 (1999).
\bibitem{Jarlskog}C.~Jarlskog, 
  Phys. Rev. Let. {\bf 55} 1039 (1985).
\bibitem{E391a}Search for CP violating decay $K_L\to\pi^0\nu\nu$:\\ 
  {\tt http://psux1.kek.jp/\~{}e391/};
  T.~Inagaki, in KAON2001.
\bibitem{JHF}High Intensity Proton Accelerator Facility:\\ 
  {\tt http://jkj.tokai.jaeri.go.jp/}.
\bibitem{KOPIO}KOPIO $K^0_L\to\pi^0\nu\nu$ Experiment:\\ 
  {\tt http://pubweb.bnl.gov/users/e926/www/index.html};
   D.~Bryman, in KAON2001.
\bibitem{CKM}CKM (E921) WWW Server:\\ 
  {\tt http://www.fnal.gov/projects/ckm/Welcome.html}; 
  E.~Ramberg, in KAON2001. 
\bibitem{Sanda}
  I.I.~Bigi and A.I.~Sanda, 
  CP violation (Cambridge University Press, Cambridge, 2000).
\bibitem{E246pub}M.~Abe {\it et al}, 
  Phys. Rev. Lett. {\bf 83}, 4253 (1999).
\bibitem{Imazato}J.~Imazato,
  in KAON2001. 
\bibitem{E246form}S.~Shimizu {\it et al}, 
  Phys. Lett. B {\bf 495}, 33 (2000).
\bibitem{E246ratio}K.~Horie {\it et al}, 
  Phys. Lett. B {\bf 513}, 311 (2001).
\bibitem{PSI}MEG Home Page:
  {\tt http:http://meg.psi.ch/}.
\bibitem{MECO}Muon to Electron COoversion Experiment (MECO):\\
  {\tt http://meco.ps.uci.edu/}.
\bibitem{Langacker}P.~Langacker {\it et al}, 
  Phys. Rev. D {\bf 38}, 2841 (1988).
\bibitem{SUSY}B.A.~Campbell,
  Phys. Rev. D {\bf 28}, 209 (1983).
\bibitem{CHS}R.N.~Cahn and H.~Harari, 
  Nucl. Phys. B {\bf 176}, 135 (1980); 
  O.~Shanker, 
  Nucl. Phys. B {\bf 206}, 253 (1982). 
\bibitem{E871mue}D.~Ambrose {\it et al}, 
  Phys. Rev. Lett. {\bf 81}, 5734 (1998).
\bibitem{KTEVpimue}A.~Bellavance,
  in the DPF2000 meeting (Ohio, August 2000). 
\bibitem{E865pimue}R.~Appel {\it et al}, 
  Phys. Rev. Lett. {\bf 85}, 2450 (2000).
\bibitem{E777}A.M.~Lee {\it et al}, 
  Phys. Rev. Lett. {\bf 64}, 165 (1990).
\bibitem{E865db}R.~Appel {\it et al}, 
  Phys. Rev. Lett. {\bf 85}, 2877 (2000).
\bibitem{PDG86}Particle Data Group, M.~Aguilar-Benitez {\it et al}, 
  Phys. Lett. B {\bf 170}, 1 (1986).
\bibitem{PDG96}Particle Data Group, R.M.~Barnett {\it et al}, 
  Phys. Rev. D {\bf 54}, 1 (1996).


\end{thebibliography}
\end{document}